\documentclass[12pt]{article}

\usepackage{amssymb,a4wide}
\usepackage{epsfig}
\usepackage{pictex}
\usepackage[usenames]{color}
\usepackage{subfigure}
\usepackage{rotating}
\usepackage{graphics}
\usepackage{fancyheadings}
\usepackage{cite}
\usepackage{amsmath}
\usepackage{amssymb}

\def\be{\begin{equation}}
\def\ee{\end{equation}}
\def\C{\mathcal{C}}

\def\hr{\hat{\rho}}
\def\hP{\hat{P}}

\begin{document}

\title{{\huge Graviton Emission into Non-$Z_2$ Symmetric Brane World Spacetimes.}}

\author{
Ian R. Vernon\footnote{I.R.Vernon@dur.ac.uk},
David Jennings \footnote{D.Jennings@damtp.cam.ac.uk},\\
\normalsize \em Centre for Particle Theory, University of Durham,\\
\normalsize \em South Road, Durham, DH1 3LE, U.K.\\
\normalsize \em Department of Applied Mathematics and Theoretical Physics,\\
\normalsize \em CMS, University of Cambridge, Cambridge, CB3 0WA, U.K.\\
}

\maketitle

\begin{abstract}

The equations for the evolution of a homogeneous brane world that emits gravitons at 
early times, and into a non-$Z_2$ symmetric bulk, are derived using an 
$AdS$-Vaidya spacetime approximation. The behaviour of the black hole mass parameters either 
side of the brane is analysed, and it is found that in general graviton emission leads 
to a decrease in the non-$Z_2$ symmetry. However, the behaviour of the dark radiation 
term in the Friedmann equation is more complex: it is shown that this term can increase or decrease due to the 
non-$Z_2$ symmetry, and can become negative in some cases, leading to $H=0$ and the brane universe 
collapsing. Constraints on the initial (nonzero) sizes of the mass parameters are therefore derived.

\end{abstract}


\newpage

\section{Introduction}

Recently there has been considerable interest in the novel suggestion
that we live in a Universe that possesses more than four
dimensions. The standard model fields  
are assumed to be confined to a hyper-surface (or 3-brane)
embedded in this higher dimensional space, in contrast to the gravitational
fields which propagate through the whole of
spacetime~\cite{ruba,akam1,anton,visser,ahdd,ahddII,aahdd,sund1,rscompact99,rshierarchy99}.
In order for this to be a
phenomenologically relevant model of our universe, 
standard four-dimensional gravity must be recovered on our
brane. There are various ways to do this, the most obvious being to
assume that the extra dimensions transverse to our brane are compact.
In this case gravity can be recovered on scales larger than the size
of the extra dimensions~\cite{ahdd,ahddII,aahdd}. This is different
from earlier proposals since the restrictions on the size of the 
extra dimensions from particle
physics experiments no longer apply, as the standard model fields 
are confined to the brane. The extra dimensions only have to
be smaller than the scale on which gravity experiments have probed,
of order 0.1mm at the time of writing.
Another way to recover four-dimensional gravity at large distances is 
to embed a positive tension 3-brane into an
AdS$_5$ bulk~\cite{rscompact99,rshierarchy99}. In this scenario four-dimensional
gravity is obtained at scales larger than the AdS radius. Randall
and Sundrum showed that this could produce sensible gravity even if
the extra dimension was not compact.

Several interesting aspects of the above extra dimensional scenarios have since 
been investigated, and compared with the standard four-dimensional case.
The cosmology of a 3-brane in a five-dimensional bulk was studied and 
its Friedmann equation derived and shown to contain several extra 
terms~\cite{noncosmo,bulkcosmo,cosmo1extrad,expans1extrad,Kraus}. Perturbations 
to this homogeneous case have been examined~\cite{pert1,pert2,pert3,pert4} as 
have some inflationary models~\cite{inflation,infl2}, as well as alternative solutions to 
the flatness and horizon problems~\cite{ver3}. The behaviour of phase transitions, 
topological defects and baryogenesis in a brane world setting were also considered~\cite{cosmophase}. 
More recently it was shown how to embed the Randall-Sundrum models within 
supergravities~\cite{s49,s50,s51} and then within string theory compactifications 
as in~\cite{s52,s53,s54,s55}. The seemingly arbitrary feature of having an $AdS$ bulk 
spacetime is actually well motivated as it is found as a supersymmetric vacuum to 
supergravity theories, inspiring several more recent brane world 
models~\cite{anne1,anne2,anne3,anne4,anne5,anne6,anne7}. For recent, comprehensive 
reviews of the subject see~\cite{carrev,royrev}. 

The extra term $\C/r^4$, in the Friedmann equation known as the dark radiation term, could provide 
a possible test for the above extra-dimensional scenarios. Although many authors have assumed 
that $\C$ is constant, this will not be the case if the brane emits gravitons at early times. 
The evolution of $\C$ has been studied~\cite{bulkbh,lang,Leeper} for symmetric branes. However, in 
this paper we examine the dynamics of a non-$Z_2$ symmetric graviton emitting brane~\cite{Leeper2} and determine what consequences the lack of symmetry has on the evolution of $\C$.
We begin in section~\ref{NSB} by summarising the main aspects of non-$Z_2$ branes, including 
the new Friedmann equation and the global nature of the five-dimensional 
spacetimes either side of the brane. Then in section~\ref{GEVS} we derive the equations 
that govern the evolution of the 
energy density $\rho$, the dark radiation parameter $\C$ and the non-$Z_2$ symmetric parameter $F$, 
using a Vaidya spacetime approximation. The global nature of these five-dimensional 
($AdS$)-Vaidya spacetimes is then discussed along with the appropriate conformal diagrams. We then 
analytically solve the system of equations in the high/low energy limits in section~\ref{ana}, 
before presenting a full numerical treatment in section~\ref{NA}. We examine the full range 
of initial conditions for $\C$ and $F$, and include several interesting cases that have so far 
been ignored. Constraints on the size of the initial dark radiation and non-$Z_2$ symmetric 
terms in this scenario are derived and are notably independent of $M_5$, the five dimensional 
Plank mass. Our conclusions are given in section~\ref{conc}.

\section{Non-$Z_2$ Symmetric Branes Moving Through Five-dimensional Spacetimes}\label{NSB}

In this section we summarise the relevant 
aspects of a non-$Z_2$ brane in an $AdS$ bulk. We discuss the derivation of the 
non-$Z_2$ Friedmann equation and also the global nature of the 
five-dimensional Schwarzschild-Anti-deSitter spacetime.

\subsection{The Non-$Z_2$ Symmetric Friedmann Equation}
Starting with the five-dimensional Einstein equations,
\be\label{ein}
G_{AB} \;\;+\;\; \Lambda g_{AB} \;\;\;=\;\;\; \kappa^2 T_{AB},
\ee
where $\kappa$ is related to the fundamental five-dimensional Plank 
mass $M_5$ by $\kappa = 1/M_5^3$, 
one can use Birkhoff's theorem to show that the bulk metric must take the form,
\be\label{met}
ds^2 \;\;=\;\; - f^{\pm}(r) dT^2 \;\;+\;\;r^2 d\Sigma_k^2 \;\;+\;\; \frac{1}{f^{\pm}(r)} dr^2.
\ee
Here $d\Sigma_k^2$ represents the maximally symmetric three-dimensional metric with $k=-1,0,1$ 
parameterising the spatial curvature. The $\pm$ in equation~\ref{met} represents the fact that 
the bulk spacetime can be different either side of the brane. 
The function $f^{\pm}(r)$ is found by substituting the 
metric~(\ref{met}) into the Einstein equations~(\ref{ein}) and is given by~\cite{Kraus},
\be\label{fpm}
f^{\pm}(r) \;\;=\;\; \mu^2 r^2 \;+\; k \;-\; \frac{\C^{\pm}}{r^2}.
\ee
$\mu$ is the inverse curvature radius, related to the five-dimensional cosmological constant by 
$\mu = \sqrt{-\Lambda/6}$, and $\C^{\pm}$ is the bulk black hole mass parameter to the 
right/left of the brane respectively. We see that the bulk solution on either side of the brane can be either an 
$AdS$ black hole $(\C^{\pm}>0)$, an $AdS$ naked singularity $(\C^{\pm}<0)$ or pure $AdS$ $(\C^{\pm}=0)$. 

If a brane with stress-energy-momentum tensor 
$T^A{}_B = \delta(r) {\rm diag}(-\hr -\sigma, \hP -\sigma, \hP -\sigma,\hP -\sigma,0)$ 
exists between two such spacetimes, the Israel junction conditions~\cite{s22,s23,s26,s27} 
can be used to show that 
the Friedmann equation governing the expansion of the brane is given by~\cite{fpaper},
\begin{equation}\label{fri}
\hat{H}^2 \;\; \equiv \;\; \left( 
      \frac{\dot{r}}{r}
\right)^2 = 
\frac{\kappa^4 \sigma}{18} \hr  \; + \; 
\frac{\kappa^4}{36}\hr^2 \; - \; \frac{k}{r^2} \; + \; 
\frac{\mu^2 \C}{r^4} \; + \; \frac{\mu^2 \sigma^2 F^2}{(\hr + \sigma)^2 r^8}.
\end{equation}
Here $\sigma$ is the bare brane tension, $\rho$ is the physical energy density of the brane, and 
we have imposed the Randall-Sundrum tuning conditions such that 
$\kappa^4 \sigma^2/36 + \Lambda/6 = 0$, and hence that $\kappa^4 \sigma^2/36 = \mu^2$. 
The Weyl (or dark radiation) parameter $\C$ and the 
non-$Z_2$ parameter $F$ are dimensionless, and are related to the average and difference 
of the black hole masses either side of the brane via,
\be\label{CF}
\C \;\;=\;\; \frac{\C^+ + \C^-}{2\mu^2}, \;\;\;\;\; F \;\;=\;\; \frac{\C^+ - \C^-}{4\mu^2}.
\ee
The implications of the extra terms in the Friedmann equation~(\ref{fri}) have been 
thoroughly investigated~\cite{noncosmo,bulkcosmo,fpaper,ver4}. However, 
as will be discussed in the following section, if gravitons 
are emitted by the brane into the bulk at early times then the parameters $\C$ and 
$F$ will no longer be constant and the early cosmological evolution of the brane 
could be greatly 
affected. Note that there is an alternative method of breaking the $Z_2$ symmetry across 
the brane; one can have 
different cosmological constants either side of the brane leading to a term 
$\propto 1/r^6$ in the 
Friedmann equation. Although interesting, this type of non-$Z_2$ symmetry is not directly 
affected by graviton emission and so will not be considered here.

\subsection{Global Structure of Sch-AdS}

Here we summarise some aspects of the global structure of the aforementioned Sch-AdS spacetime 
and the brane's trajectory through it. For clarity we refer only to the spacetime on the right hand side of the brane, 
which has black hole mass $\C^+$. Such concepts will be of use when in the next section we 
generalise the analysis to Vaidya spacetimes. In what follows, we assume the case where 
$k=0$ and $\C^+$ is positive. Other cases can be argued similarly.

One should note that the coordinates $T$ and $r$ in equations~(\ref{met}) and (\ref{fpm}) do not 
cover the whole of the Sch-AdS spacetime, and are valid only for $r > r_h = (\C^+)^{1/4}/\mu^{1/2}$. 
To continue beyond this horizon we first note that the metric given by equation~(\ref{met}) 
is conformally equivalent (after rescaling) to,
\be
ds^2 \;\; = \;\; -\left(
                       \frac{r^4 - 1}{r^2}
                  \right) \: d T^2 \;+\; 
                  \left(
                       \frac{r^2}{r^4 - 1}
                  \right) \: d r^2 \;+\; 
                  r^2 d\Sigma_0^2,
\ee
where the horizon is now at $r=1$. We then map the radial 
coordinate from $1<r<\infty$ to $-\infty < R < \pi/4$ according to the transformation,
\begin{equation}
R \;\;=\;\; \int\; \frac{r^2 dr}{r^4-1} \;\; =\;\; 
            \frac{1}{2} \tan^{-1}(r) \;+\; 
            \frac{1}{4}\log \left[\frac{r-1}{r+1} \right].
\end{equation}
Then after changing to the null coordinates given by $u=t-R$ and $v=t+R$ 
the metric looks like,
\begin{equation}
ds^2 \;\; = \;\; - \left( \frac{r^4 -1}{r^2} \right) \; du dv \;+\; 
                   r^2 d\Sigma_0^2.
\end{equation}
Finally changing to the coordinates $U$ and $V$ given by,
\begin{eqnarray}
U & = & -\exp(-2u),\\
V & = & \exp(2v),
\end{eqnarray}
results in the metric,
\begin{equation}
ds^2 \;\;=\;\; -\left( \frac{r^4-1}{r^2} \right)
                \left( \frac{dU dV}{-4UV} \right) \;+\; r^2 d \Sigma_0^2.
\end{equation}
Note that the point $(1,0)$ in the $(r,t)$ plane has now been mapped onto 
$(0,0)$ in the $(U,V)$ plane and hence that the future and past horizons at 
$r=1$ of the black hole have been mapped onto the positive $V$ and negative $U$ 
axes respectively. To see that the zero at $r=1$ has been removed note that,
\begin{equation}
U V \;\;= \;\; - \left(\frac{r-1}{r+1} \right) \exp[2 \tan^{-1}(r)],
\end{equation}
which behaves as $(r-1)$ for $r \approx 1$. 
\begin{figure}[hbtp]
       \centerline{     
        \scalebox{0.5}{
        \input{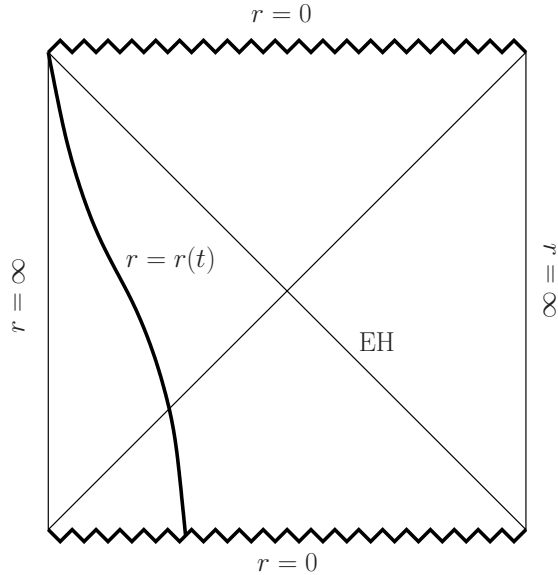} 
          }
        }
      \caption{Conformal diagram of the extended Schwarzschild Anti-DeSitter spacetime. The thick 
curved line represents the trajectory of a cosmologically realistic brane, while EH labels the
event horizon. Note that here, and in all subsequent figures only the spacetime to the right of the brane exists.}
     \label{SchAdS}
\end{figure}
Using the above coordinate transformation one can deduce that the conformal diagram of 
five-dimensional Sch-$AdS$ is given by figure~\ref{SchAdS}. The left and right vertical 
boundaries represent the spatial infinities, while the top and bottom boundaries correspond 
to the black and white hole singularities respectively. Although the Friedmann 
equation~(\ref{fri}) is derived from the metric~(\ref{met}), it is still valid in all 
regions of the Sch-$AdS$ spacetime, and hence we can use it to examine the trajectory 
of a brane moving through such a spacetime. A cosmologically realistic brane that starts
from an initial singularity $r=0$ with $\C^+>0$, and then expands forever will have the trajectory 
shown in figure~\ref{SchAdS}. It begins at the white hole singularity, passes 
through the event horizon at $r=1$ and continues off to infinity. 

We can ask how the situation would change if the brane is emitting gravitons toward the 
bulk, and more specifically how this would effect the black/white hole 
mass parameter $\C^+$. In the next section we use a five-dimensional $AdS$ 
generalisation of the ingoing Vaidya spacetime to model the effect of graviton emission 
by the brane, and hence derive the evolution equations for $\C^+$ and $\C^-$. We then go 
on to discuss the extension of Vaidya spacetimes, and present suitable conformal diagrams 
to explain conceptually the results given in sections~\ref{ana} and \ref{NA}.

\section{Graviton Emission and Vaidya Spacetimes}\label{GEVS}

In the previous section we discussed a brane moving through two different spacetimes possessing 
only a cosmological constant. We now go on to generalise the situation presented by 
Langlois et.al.~\cite{lang}, where the brane emits 
gravitons at early times, to the above non-$Z_2$ scenario~\cite{Leeper2}. 

In order to model this situation we use the five-dimensional generalisation of the 
incoming\footnote{In the bulk the brane resides at the largest radius of the spacetime, hence 
any gravitons emitted will travel inward.}
Vaidya metric,
\be\label{vad}
ds^2 \;\;=\;\; -f^{\pm}(r,v) dv^2 \;+\; 2drdv \;+\; r^2 d\Sigma_k^2,
\ee
where we once again assume a different spacetime either side of the brane and that $k=0$,
\be
f^{\pm}(r,v)\;\;=\;\; \mu^2 r^2 - \frac{\C^{\pm}(v)}{r^2}.
\ee
Here $v$ is a null ingoing coordinate. Note that now the Weyl parameters $\C^{\pm}(v)$ 
no longer represent purely the mass of the black hole in each spacetime, but instead 
represent the energy contained inside the radius $r$ for a given $v$, and hence 
describes an inward flow of radiation. If $\C^{\pm}$ did not depend on $v$, then the 
metric~(\ref{vad}) would just be a rewriting of Schwarzschild-AdS, 
given by equation~(\ref{met}), as can be seen by the coordinate transformation 
$v = T + \int dr / f(r)$.
The brane moves through both such spacetimes with trajectories given by 
$v^{\pm}(\hat{t})$ and $r(\hat{t})$, where $\hat{t}$ is the time experienced by a brane based observer. 
Normalisation of the brane's velocity vector in each spacetime 
$u^{A\pm} = \{ \dot{v}^{\pm}, \dot{r},0,0,0 \} $ requires that,
\be\label{vdot}
\dot{v}^{\pm} \;\;=\;\; \frac{\dot{r} + \sqrt{f^{\pm} + \dot{r}^2}}{f^{\pm}},
\ee
which we will need later.

The incoming Vaidya metric is a solution to Einstein's equations with a bulk 
energy-momentum-tensor corresponding to null ingoing radiation,
\be\label{set}
T_{AB}^{\pm} \;\;=\;\; \psi k_A^{\pm} k_B^{\pm},
\ee
where $k^{\pm A}$ are null ingoing vectors in each spacetime. By inserting the metric~(\ref{vad}) 
and the stress-energy-tensor~(\ref{set}) into Einstein's equations we find that they 
will be solved provided the Weyl parameters satisfy,
\be\label{Cv}
\frac{d \C^{\pm}(v)}{d v} \;\;= \;\; \frac{2\kappa^2 \psi r^3}{3} \: k_v^{\pm} k_v^{\pm}.
\ee
The appropriate normalisation of $k^{\pm A}$ is given by $k^{\pm}_A u^{\pm A} = 1$, which implies that the only 
non-zero component is $k^{\pm r} = k^{\pm}_v = 1/\dot{v}^{\pm}$. Also $\psi $ is the flux of gravitons 
leaving a radiation dominated brane as seen by a brane observer, and has been shown by 
Langlois et. al.~\cite{lang} to be,
\be\label{psi}
\psi \;\;=\;\; \frac{\alpha}{12} \kappa^2 \hat{\rho}^2,
\ee
where $\alpha$ is a dimensionless constant. Note that here we implicitly assume that gravitons are 
produced in equal amounts either to the left or right of the brane. Particle interactions 
that produce these gravitons will not feel the effect of different bulk masses either side 
of the brane. However we will go on to show that the difference in the brane's trajectory in 
each spacetime alters the effect of the graviton emission so that it appears to be non-$Z_2$ 
symmetric. 

Combining equations~(\ref{vdot}), (\ref{Cv}) and (\ref{psi}), we get an expression for 
the evolution of either Weyl parameter due to graviton emission,
\be\label{wpm}
\frac{d \C^{\pm}}{d \hat{t}} \;\;=\;\; \frac{\alpha \kappa^4}{18} r^3 \hat{\rho}^2 
                           ( \sqrt{ \dot{r}^2 + f^{\pm}} - \dot{r}).
\ee
In order to determine the behaviour of a brane positioned between two such Vaidya spacetimes, 
we need to impose the Israel junction conditions~\cite{s22,s23,s26,s27} across the brane. 
These conditions relate the jump in the extrinsic curvature of the brane 
$K_{AB} = h^C{}_A \nabla_C n_B$, where as usual $n^A$ is the unit vector normal to the brane 
and $h_{AB} = g_{AB} - n_A n_B$ the induced metric, to the brane's energy momentum tensor 
$\tau_{AB} = {\rm diag}(-\hr -\sigma, \hP -\sigma, \hP -\sigma,\hP -\sigma,0)$. This can 
be written in the covariant form:
\be\label{JC}
\left[ K_{AB} \right] \;\;=\;\; \kappa^2 (\tau_{AB} \;-\; \frac{1}{3} \tau h_{AB}).
\ee
The ordinary spatial components of these Junction Conditions show that the Friedmann equation 
for the brane is of the same form as that of equation~(\ref{fri}). However, now both the 
Weyl parameter $\C$ and the non-$Z_2$ parameter $F$ are now functions of $v$ and hence $t$. 
Using the fact that $H=\dot{r}/r$, we can use equations~(\ref{fri}) and (\ref{fpm}) to 
rewrite equation~(\ref{wpm}) as,
\be\label{cpm2}
\frac{d \C^{\pm}}{d \hat{t}}\;\; = \;\; 
            \frac{\alpha \kappa^4 \hr^2 r^4}{18}
            \left(
                  \frac{\kappa^2}{6}(\hr + \sigma) \: \mp \: 
                   \frac{3}{2\kappa^2} \frac{(\C^+ - \C^-)}{(\hr + \sigma)r^4} \:-\; H
           \right).
\ee
It can immediately be seen that the process of graviton emission will lead to a reduction of 
the $Z_2$ symmetry since if $\C^+$ is greater/less than $\C^-$ then $d\C^+/dt$ will 
be less/greater than $d\C^-/dt$, and hence the non-$Z_2$ parameter $F$ will decrease. To 
analyse this in more detail we take the sum and difference of equations~(\ref{cpm2}) to 
obtain expressions for the evolution of the Weyl parameter $\C$ and the non-$Z_2$ symmetry 
parameter $F$ defined in equations~(\ref{CF}),
\begin{eqnarray}\label{dC}
\frac{d\C}{d \hat{t}} &=& \frac{\alpha \kappa^4 \hr^2 r^4}{18\mu^2}
                   \left( 
                       \frac{\kappa^2}{6}(\hr + \sigma) \:-\: H
                  \right),  \\ \label{dF}
\frac{d F}{d \hat{t}} &=& - \frac{\alpha \kappa^2 \hr^2}{6(\hr + \sigma)} F.
\end{eqnarray}
These equations will be examined both analytically and numerically in sections~\ref{ana} and \ref{NA} 
for various cases of interest.

The time and fifth dimension part of the Junction Conditions, given by equation~(\ref{JC}), 
also yield another relation representing the non-conservation of energy on the brane,
\be\label{noncon}
\dot{\hr} \;+\; 3\frac{\dot{r}}{r} (\hr + \hP) \;\;=\;\; -2 \psi.
\ee
The factor of 2 on the right hand side is due to the fact that the brane is emitting 
a flux $\psi$ of gravitons into each spacetime either side of the brane.
We have hence derived the four `master' equations that describe brane graviton emission into 
non-$Z_2$ symmetric spacetimes, given by~(\ref{fri}), (\ref{dC}), (\ref{dF}) and 
(\ref{noncon}). In the next section we investigate the global properties of Vaidya spacetimes in order to 
understand the results presented in sections~\ref{ana} and \ref{NA}.

\subsection{Extension of the 5D Vaidya-AdS Spacetime}\label{ev}

We have derived the relevant equations that describe both the motion of a graviton 
emitting non-$Z_2$ symmetric brane and the evolution of the mass parameters either side 
of the brane. However, one should note that the coordinates $(v,r)$ do not cover the 
whole of the spacetime. Therefore we now show how to extend the Vaidya coordinates 
(see~\cite{fayos1,fayos2} for four dimensional examples), 
in order to gain a conceptual understanding of certain cases of interest, including 
when the brane emerges from the white hole. Previous authors have ignored these 
solutions, mainly by assumption~\cite{lang,Leeper2,mis1}.
We mainly concentrate on the situation where 
the BH mass parameter $\C^+(v)$, is positive for all time; cases where $\C^+(v)<0$ or 
where $\C^+(v)$ changes sign can again be argued similarly~\cite{fayos2}.

The ingoing five-dimensional Vaidya metric~(\ref{vad}) is conformally equivalent 
(after rescaling $r\rightarrow r/\mu$ and $v\rightarrow v/\mu$) to:
\be\label{V1}
ds^2 \;\;=\;\; -\left(
                r^2 - \frac{r_h^4(v)}{r^2}
               \right) dv^2 \:+\: 2 dv dr \:+\: r^2 d\Sigma_0^2,
\ee
where $r_h^4(v) = \mu^2 \C^+(v)$. Examining the Friedmann equation~(\ref{fri}), it can be seen 
that if the BH mass parameter is initially 
greater than zero ($\C^+(v \rightarrow -\infty)>0$), the brane will originate from 
the singularity inside the white hole ($r=0$) and emerge from it at a later time,
similar to the non-emitting case. The above Vaidya coordinates only cover the interior 
region of the black hole (along with one of the exterior regions) and hence need to 
be extended. 
\begin{figure}[h]
       \centerline{     
        \scalebox{0.5}{
        \input{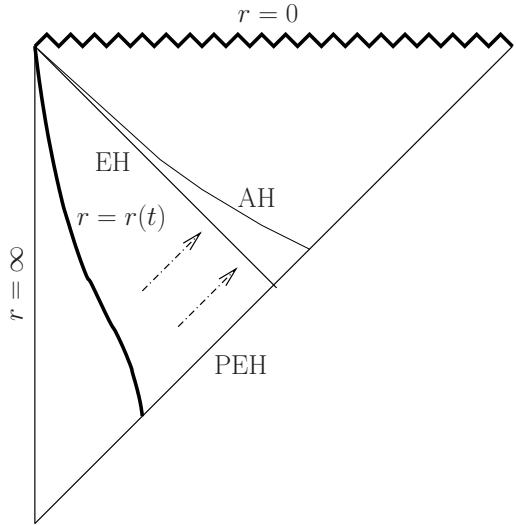} 
          }
        }
      \caption{Conformal diagram of the unextended ingoing $AdS$ Vaidya spacetime}
     \label{Vaid2}
\end{figure}
The conformal diagram 
of the region covered by the $(v,r)$ coordinates is shown in figure~\ref{Vaid2}, along 
with a possible trajectory of the brane ($r=a(t)$), the direction of graviton 
emission (the dashed arrows) and the various horizons. EH represents the normal 
event horizon of the black hole, whether PEH represents the past event horizon 
appearing at $r=r_h(v\rightarrow - \infty)$. It can be seen that outgoing radial null 
geodesics emerge from the PEH at finite values of their affine parameter and are hence 
incomplete: extension is therefore required. 

Figure~\ref{Vaid2} also shows one of the main features of non-stationary spherically 
symmetric spacetimes, which is the decoupling of the apparent horizon (AH) from the 
event horizon~\cite{fayos1}. The apparent horizon is the hypersurface separating the regions with or 
without closed trapped surfaces (see~\cite{ellis} for more details), here given 
by $r=r_h(v)$, and will be matched to a corresponding AH in the extended spacetime.

In general, extending Vaidya spacetimes is non-trivial as is discussed by~\cite{fayos1}. 
One usually does not know the form of the `mass function' $\C^+(v)$ 
(and therefore $r_h(v)$) in the extended part 
of the spacetime or indeed even whether radiation is ingoing or outgoing there. This 
leads to many possible choices of extension.
However, in the case whereby a brane emits the radiation into the bulk, we do in principle 
know the general form of $\C^+(v)$, and hence we can adapt Israel's extension of 
four-dimensional Vaidya spacetimes~\cite{israel} to the five-dimensional $AdS$ Vaidya metric 
given in equation~(\ref{V1}). 

Changing coordinates from $(v,r)$ to $(U,V)$ by the transformation:
\be
r \;\;=\;\; A U \:+\: r_h , \quad \quad v \;\;=\;\; \int^V \frac{d x}{A(x)},
\ee
where the function $A(V)$ is chosen to be:
\be 
A(V) \;\;=\;\; \int^V_0 2 r_h(x) dx,
\ee
the metric given by equation~(\ref{V1}) then becomes,
\be 
ds^2 \;\;\;=\;\;\; \left( 
           \frac{U^2(-A^2 U^2 + 2 r_h^2)}{(AU + r_h)^2} \; +\;  \frac{2}{A} \:\frac{d r_h}{d V}
                  \right) \: dV^2 \,+\,2 \:dU dV \,+\, (AU + r_h)^2 \:d\Sigma_0^2.
\ee
Note that the PEH has been mapped from $(r=r_h,v=-\infty)$ to $(-\infty<U<\infty, V=0)$, 
and AH from $(r=r_h(v), -\infty <v<\infty )$ to $(U=0, 0< V< \infty)$. The extended 
part of the spacetime corresponds to $-\infty <V< 0 $. Some conditions must be imposed 
upon the mass function $r_h(V)$ to ensure the finiteness and continuity of both the metric 
and the energy momentum tensor~\cite{israel} 
(which we know to still describe radiation leaving the brane), which are that 
$r_h(V)$ must be a $C^2$ function such that~\cite{fayos1},
\be 
r_h(V=0) \; \ne \; 0, \quad \quad  
\left. \frac{d r_h}{d V} \frac{1}{A}  \right|_{V=0} \; < \; \infty \;\;\Rightarrow \;\;
\frac{d r_h}{d V} (V=0) \:=\:0. 
\ee
We have assumed the first condition to be true previously. As for the second condition,
demanding that the full metric satisfies the energy conditions~\cite{israel} implies 
that,
\be
\frac{d r_h}{d V} \;\;\; \left\{ 
 \begin{array}{cc}
 \ge \: 0 &\; {\rm for}\;\;\; V>0 \\
 \le \: 0 &\; {\rm for}\;\;\; V<0 \\
\end{array}\right.
\ee
and hence the second condition is satisfied purely by demanding that the metric is 
physically realistic. 
This has some interesting implications:
we can see that when the brane is in the extended part ($V<0$) of the spacetime, 
any graviton emissions will actually decrease the size of the mass function $r_h(V)$, 
and hence decrease $\C^+(V)$. 
\begin{figure}[hbtp]
       \centerline{     
          \scalebox{0.5}{
         \input{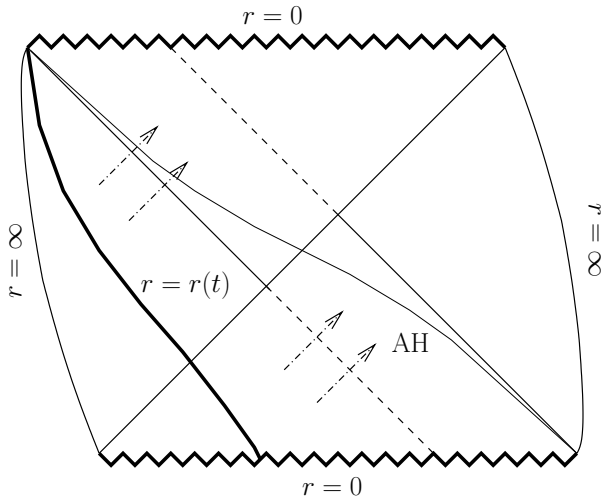} 
            }
        }
     \caption{Conformal diagram of the extended ingoing $AdS$ Vaidya spacetime}
     \label{Vaid1}
\end{figure}
The conformal diagram for the extended spacetime is 
given by figure~\ref{Vaid1}, which also shows the the direction of the emitted gravitons and the 
trajectory of the brane. This trajectory is a possible solution $r(t)$ to the 
equations~(\ref{fri}), (\ref{dC}), (\ref{dF}) and (\ref{noncon}) which are still valid in the extended 
part of the spacetime. One can see that the 
$V<0$ region is actually a region of outgoing Vaidya spacetime with the metric,
\be\label{V2}
ds^2 \;\;=\;\; -\left(
                \tilde{r}^2 - \frac{\tilde{r}_h^4(\tilde{v})}{\tilde{r}^2}
               \right) d\tilde{v}^2 \:-\: 2 d\tilde{v} d\tilde{r} \:+\: \tilde{r}^2 d\Sigma_0^2.
\ee
One should also note that the coordinates $(U,V)$ used to extend the spacetime are 
only valid provided $AU < \sqrt{2} r_h$, which means they cover the interior of both 
the black and white holes, but do not extend to spacelike infinity ($r=\infty$).

Figure~\ref{Vaid1} shows that any gravitons emitted before the brane crosses $V=0$ (equivalent to $r=r_h$) 
are sent in an outward null direction and hence leave the white hole,  
subsequently escaping to infinity, and will therefore cause the mass parameter $\C^+$ to decrease. 
This is in agreement with equation~(\ref{cpm2}) which gives that $d\C^+/dt$ will 
be greater/less than zero if $r$ is greater/less than $r_h$ (and a similar result for $\C^-$).
This situation where the mass parameter initially decreases will occur at early times in the 
brane's cosmological evolution, when the brane will be at high temperature. 
This raises the possibility that graviton emission at early times 
could help to decrease $\C^+$ (and/or $\C^-$ and therefore also $\C$), 
and hence decrease the amount of dark radiation 
below the bounds set by the abundance of light elements at 
nucleosynthesis. Figure~\ref{Vaid1} shows the case where the mass function is 
symmetric either side of $V=0$, but for an expanding, cooling brane thermally emitting 
gravitons the situation would be very asymmetric and could lead to the final mass of the 
black hole being substantially smaller than that of the white hole. 

\begin{figure}[hbtp]
       \centerline{     
          \scalebox{0.5}{
         \input{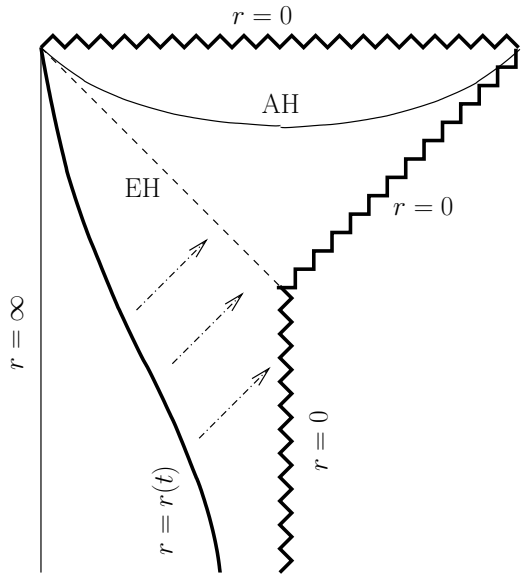} 
            }
        }
     \caption{Conformal diagram of a naked singularity changing to a black hole}
     \label{Vaid0}
\end{figure}

The case where the black hole mass parameter is initially negative, before changing 
to positive is shown in figure~\ref{Vaid0}. Initially, $\C^+<0$ and there is a naked singularity 
at $r=0$. However, provided the brane emits enough gravitons, $\C^+$ will change sign and a black 
hole will be formed~\cite{fayos2}. In this situation, equation~(\ref{cpm2}) shows that $\C^+$ always increases. 
One must of course note that a naked singularity in the bulk is phenomenologically problematic for a realistic 
brane world model.
We can now use figures~\ref{Vaid1} and \ref{Vaid0} to understand the general case. As has been discussed 
above, regardless of the initial values of $\C^+$ and $\C^-$, the non-$Z_2$ parameter $F$ will always decrease. 
The behaviour of the dark radiation parameter $\C$ is as follows: if $0<\C^- < \C^+$ (the spacetimes either side 
of the brane both correspond to figure~\ref{Vaid1}), then $\C$ will initially decrease until some time where 
$\C^- < \mu^2 r^4 < \C^+$, from which point $\C$ will increase forever, asymptotically approaching a constant value as 
$t\rightarrow \infty$ and the brane cools. If $\C^- < \C^+ < 0 $, (the spacetimes either side 
of the brane both correspond to figure~\ref{Vaid0}), then $\C$ will increase for all time, 
asymptotically approaching a constant as before. If $\C^- < 0 < \C^+$, (the spacetime on the right/left side of the 
brane corresponds to figure~\ref{Vaid1}/\ref{Vaid0}), then initially $\C^-$ will increase, $\C^+$ will 
decrease and the exact behaviour of $\C$ depends on the relative sizes of $\C^+$ and $\C^-$. 
In each of the above cases the final magnitude (and sign) of $\C$ is as yet undetermined so in the next two sections 
we give analytic and numerical solutions that can be understood in terms of the conceptual arguments 
presented  in this section. 

Throughout this paper we have assumed that the gravitons are emitted in a perpendicular direction from the 
brane and hence that a Vaidya spacetime is a realistic model of the situation. Examining the case 
where gravitons are emitted in all directions is far more complicated~\cite{lang3}, and is left to future work. 
However, we can use 
the above conformal diagrams to gain some understanding of the effect of such emissions: 
in figure~\ref{Vaid1}, non-perpendicular gravitons would follow timelike geodesics leaving the brane. 
Therefore, when the brane is inside the white hole, not all of the gravitons that it emits during this time 
will escape to infinity, and hence the mass parameter would not decrease as much as is described both above and 
in the following two sections.

\section{Equations for the Evolution of a Graviton Emitting non-$Z_2$ Brane}\label{ana}

We now go on to discuss analytic solutions valid in the high/low energy regimes and 
when the $\C$ and $F$ terms in the 
Friedmann equation are assumed to be subdominant. More general situations will be 
examined numerically in the next section.

First it is useful to define the following dimensionless parameters,
\be
\rho \;=\; \frac{\hr}{\sigma}, \quad \quad \quad t \;=\; \mu \hat{t}, \quad \quad \quad 
   H \;=\; \frac{\hat{H}}{\mu}.
\ee
The equations~(\ref{fri}), (\ref{dC}), (\ref{dF}) and (\ref{noncon}) can then be written 
in a somewhat simpler dimensionless form,
\begin{eqnarray}\label{dimeq}
H^2 &=& \rho^2 + 2\rho + \frac{\C}{r^4} + \frac{F^2}{(\rho+1)^2 r^8}, \label{H}\\
\frac{d \rho}{dt} &=& -4H \rho - \alpha \rho^2, \label{r}\\
\frac{d \C}{dt} &=& 2 \alpha r^4 \rho^2 (\rho + 1 - H), \label{C}\\
\frac{dF}{dt} &=& -\frac{\alpha \rho^2 }{\rho+1} F.\label{F}
\end{eqnarray}
Unfortunately, finding a general analytical solution to these equations, such as that 
found by Leeper et. al.~\cite{Leeper} for the $Z_2$ symmetric case, is very difficult and 
hence we restrict our analytical investigation to examine the early and late time limits 
only.

\subsection{The Early Time/High Energy Limit}
Assuming a high energy limit such that $\rho\gg 1$, and also that the $\C$ and $F$ terms 
in the Friedmann equation are sub-dominant, we can solve the evolution equations approximately.
Under these assumptions the Friedmann equation~(\ref{H}) takes the much simpler form of 
$H^2 \simeq \rho^2$. Combining this with equation~(\ref{r}) and integrating we obtain,
\be\label{it1}
\rho(t) \;\;=\;\; \frac{1}{(4+\alpha)t}.
\ee
Equation~(\ref{r}) also shows that $\rho = \gamma / r^{4+\alpha}$, where $\gamma$ is 
a constant, and hence that,
\be 
r(t) \;\;=\;\; [\gamma(4+\alpha) t ]^\frac{1}{4+\alpha},
\ee
as opposed to the usual $r \sim t^{1/4}$ in early brane world cosmology without graviton 
emission. To find the time dependence of the non-$Z_2$ parameter $F$ we use the fact that 
$\rho \gg 1$ and hence equation~(\ref{F}) becomes $dF/dt \simeq -\alpha \rho F$, which 
when integrated using equation~(\ref{r}) gives,
\be
F \;\;=\;\; A \rho^{\frac{\alpha}{4+\alpha}} \;\;=\;\; 
            \frac{A}{[(4+\alpha)t]^\frac{\alpha}{4+\alpha}},
\ee
where $A$ is a constant set by the initial conditions. It can be seen that although, as 
expected, $F$ decreases in this limit; it does so at a much slower rate compared to $\rho$.
What is important here is the behaviour of the $F$ term in the Friedmann equation~(\ref{H}), 
and we can use the above results to show that,
\be\label{ft}
\frac{F^2}{(\rho+1)^2 r^8} \simeq \frac{A^2}{\gamma^{8/(4+\alpha)}}.
\ee
Since the $F$ term behaves as a constant this raises the possibility of it dominating the 
Friedmann equation for a period, similar to the case without graviton emission~\cite{fpaper}. 
To find the time dependence of $\C (t)$ we need to use the following approximation of $H$,
\be
H \;\; \simeq \;\; \rho \:+\: 1 \:- \: \frac{1}{2\rho}
     \left(
           1 - \frac{\C}{r^4} - \frac{F^2}{(\rho+1)^2 r^8}
    \right) \:+\: \mathcal{O}(\rho^{-2}).
\ee
Inserting this into equation~(\ref{C}) and using equation~(\ref{ft}) we obtain,
\be
\frac{d \C(t)}{dt} \;\;=\;\; \alpha r^4 \rho 
      \left(
            1 - \frac{A^2}{\gamma^{8/(4+\alpha)}} - \frac{\C(t)}{r^4}
     \right).
\ee
The $\C(t)/r^4$ term on the right hand side should not, in general, 
be neglected as was done in~\cite{lang}. Using the above expressions for $\rho(t)$ and 
$r(t)$, this becomes,
\be\label{Ch}
\frac{d\C(t)}{dt} \;\;=\;\;  \frac{B}{t^{\frac{\alpha}{4+\alpha}}} \;-\;
              \frac{\alpha \C(t)}{(4+\alpha)t},
\ee
where $B$ is a constant given by,
\be
B \;\;=\;\; \alpha \frac{
              \gamma^{\frac{4}{4+\alpha}} - A^2/ \gamma^{\frac{4}{4+\alpha}}}
             {(4+\alpha)^{\frac{\alpha}{4+\alpha}}}.
\ee
Equation~(\ref{Ch}) has the solution,
\be\label{ctt}
\C(t) \;\;=\;\; B \left( 
                        t^{\frac{4}{4+\alpha}} + \frac{t_0}{t^{\frac{\alpha}{4+\alpha}}},
                 \right)
\ee
where $t_0$ is an integration constant. Again, the important fact here is the behaviour 
of the $\C(t)/r^4$ term in the Friedmann equation, compared to the dominant $\rho^2$ term.
This can be written as,
\be\label{ct}
\frac{\C(t)}{r^4} \;\;=\;\; \frac{\alpha}{4+\alpha} 
      (1-A^2/\gamma^{8/4+\alpha}) (1+t_0 (4+\alpha)\rho).
\ee
So in general this term is not initially constant as assumed by~\cite{lang}, but instead 
has a $\rho$ dependence, which could lead to it dominating the Friedmann equation at a later 
time, as will be discussed in the next section\footnote{The derived form of both the $\C$ and 
$F$ terms in the Friedmann equation validate the assumption that they are sub-dominant in this 
regime}. This dependence agrees with the 
exact $Z_2$ symmetric solutions found in~\cite{Leeper}. Note the effect here of the non-$Z_2$ 
symmetric term: aside from acting like a cosmological constant in the Friedmann equation 
it has a significant effect on the evolution of the Weyl parameter $\C$. For example,
if initially $\C=0$ and if $A>\gamma^{4/4+\alpha}$ then $\C$ will behave as $\C \sim -r^4$ i.e. 
will grow more negative. This corresponds to the case discussed in section~\ref{ev}, where 
$\C^-<0<\C^+$ and there is initially a naked singularity/black hole on the left/right side of the brane.
If $\C$ remains negative this could result in $H=0$ which would lead to the universe collapsing, 
and has to be investigated further.

\subsection{The Late Time/Low Energy Limit}

For the low energy/late time regime, we assume both that $\rho \ll 1$ and again that 
the $C$ and $F$ terms in the Friedmann equation are sub-dominant such that now 
$H^2 \simeq 2\rho$. Using this, equation~(\ref{r}) can now be integrated to give,
\be
\rho(t) \;\;\simeq\;\; \frac{1}{\left[
                \rho_i^{-1/2} + 2\sqrt{2}(t-t_i) \right]^2},
\ee
where $\rho(t_i) = \rho_i$, $t_i$ being the time the low energy approximation becomes valid. 
Note the usual $\rho \sim t^{-2}$ dependence of standard four-dimensional cosmology. 
Equation~(\ref{r}) also shows that $r= (\gamma /\rho)^{1/4}$, and inserting this into 
equation~(\ref{F}) and integrating, one obtains the solution for the non-$Z_2$ parameter 
in this late time regime in terms of $\rho$,
\be
F(\rho) \;\;\simeq\;\; F_i \exp \left[ -\frac{\alpha \rho_i^{3/2}}{6\sqrt{2}} 
      \left(
            1 - \left(\frac{\rho}{\rho_i} \right)^{3/2}
     \right) \right],
\ee
where $F(\rho_i) = F_i$. 
Since the energy density at which the low energy approximations become valid is given by 
$\rho_i \simeq 2$, we see that in this regime $F$ roughly 
decreases by a factor of $e^{-\alpha/3}$ which 
is small as expected since the universe is now cooler and hence emits less gravitons.
Replacing the above results for $\rho(t)$, $r(t)$ and $F(t)$ into equation~(\ref{C}) and 
integrating; we get the solution for the behaviour of $\C$ in the late time limit which 
we write in terms of $\rho$ for convenience,
\be
\C(\rho) \;\simeq\; \C_i  \:+\: 
     \frac{1}{\sqrt{2}}\alpha \gamma \rho_i^{1/2}
         \left[
               1- \left( \frac{\rho}{\rho_i} \right)^{1/2} 
        \right] \:-\: 
     \frac{1}{2}\alpha \gamma \rho_i
         \left[
               1-  \frac{\rho}{\rho_i}  
        \right],
\ee
where $\C(t_i) = \C_i$. Depending upon the assumed value of $\rho_i$, we can see that 
in this limit $\C$ roughly increases by $\alpha \gamma$, a small amount as expected. Note that
in this regime the non-$Z_2$ symmetry has no effect on the evolution of $\C$, mainly due to 
being rapidly damped away as $r$ increases. To analyse the effects of larger $F$ and $C$ terms,
and to examine what happens between the regimes, we now turn to numerical methods.

\section{Numerical Analysis}\label{NA}
\begin{figure}[htbp]
\vspace{-2cm}
  \begin{center}
    \mbox{
   \subfigure[Contributions to the Friedmann equation of $\rho^2$, $2\rho$, $\C/r^4$ and 
   $F^2/((\rho+1)^2 r^8)$.]
    {\scalebox{0.3}
      {\epsfig{file=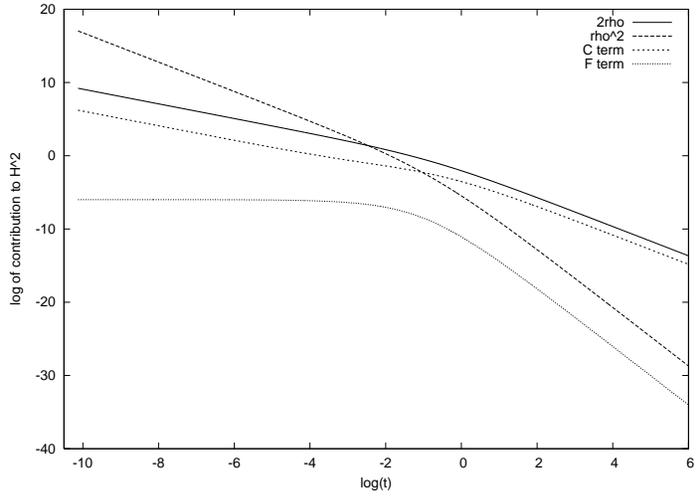,angle=180,scale=1.24}}} \quad
   \subfigure[Evolution of $\C$, $2F$, $\C^+$ and $\C^-$, with $2F_i = \C_i$ and therefore $\C_i^- = 0$.]
    {\scalebox{0.3}
      {\epsfig{file=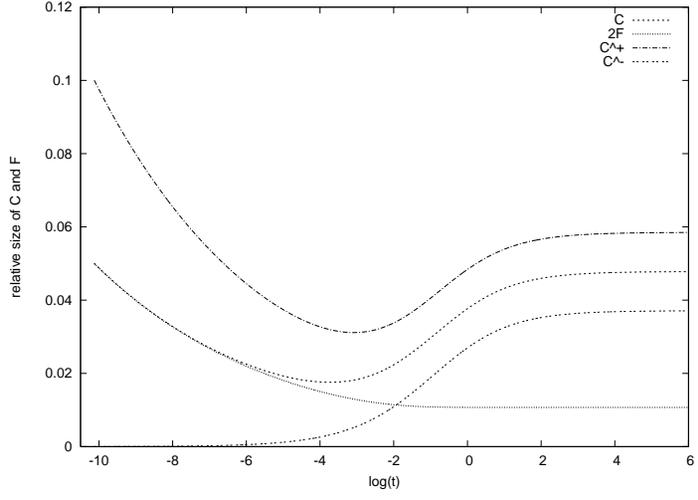,angle=180,scale=1.24}}}
      }
    \mbox{
      \subfigure[Evolution of $\C$, $2F$, $\C^+$ and $\C^-$, but initially with $\C$ and $\C^- <0$.]
    {\scalebox{0.3}
      {\epsfig{file=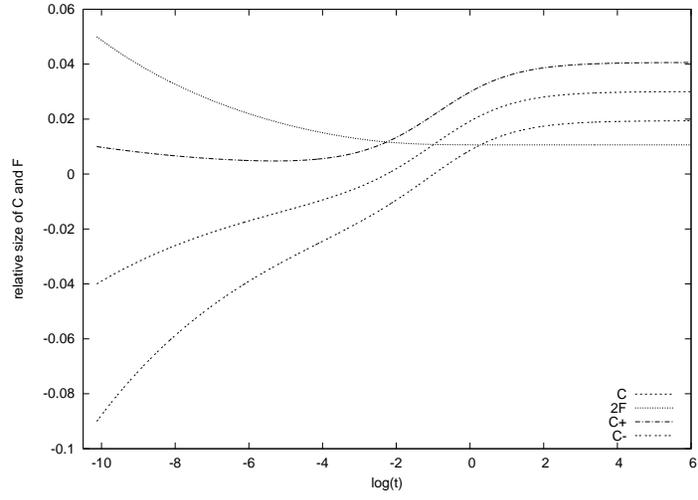,angle=180,scale=1.24}}} \quad
      \subfigure[Evolution of $\C(t)$ for different initial values of $F$.]
    {\scalebox{0.3}
      {\epsfig{file=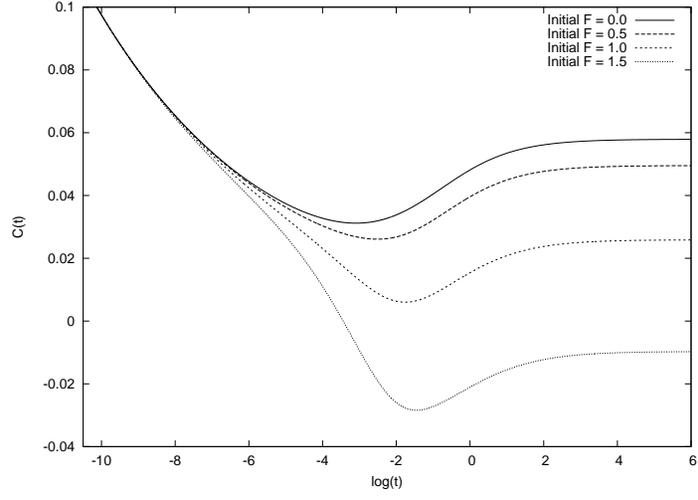,angle=180,scale=1.24}}} 
      }
    \caption{Numerical Results}\label{gra1}
  \end{center}
\end{figure}

\begin{figure}[htbp]
\vspace{-2cm}
  \begin{center}
    \mbox{
       \subfigure[Contributions to the Friedmann equation where $F$ is large enough to ensure that 
$\C$ turns negative and hence leads to $H^2=0$.]
    {\scalebox{0.3}
      {\epsfig{file=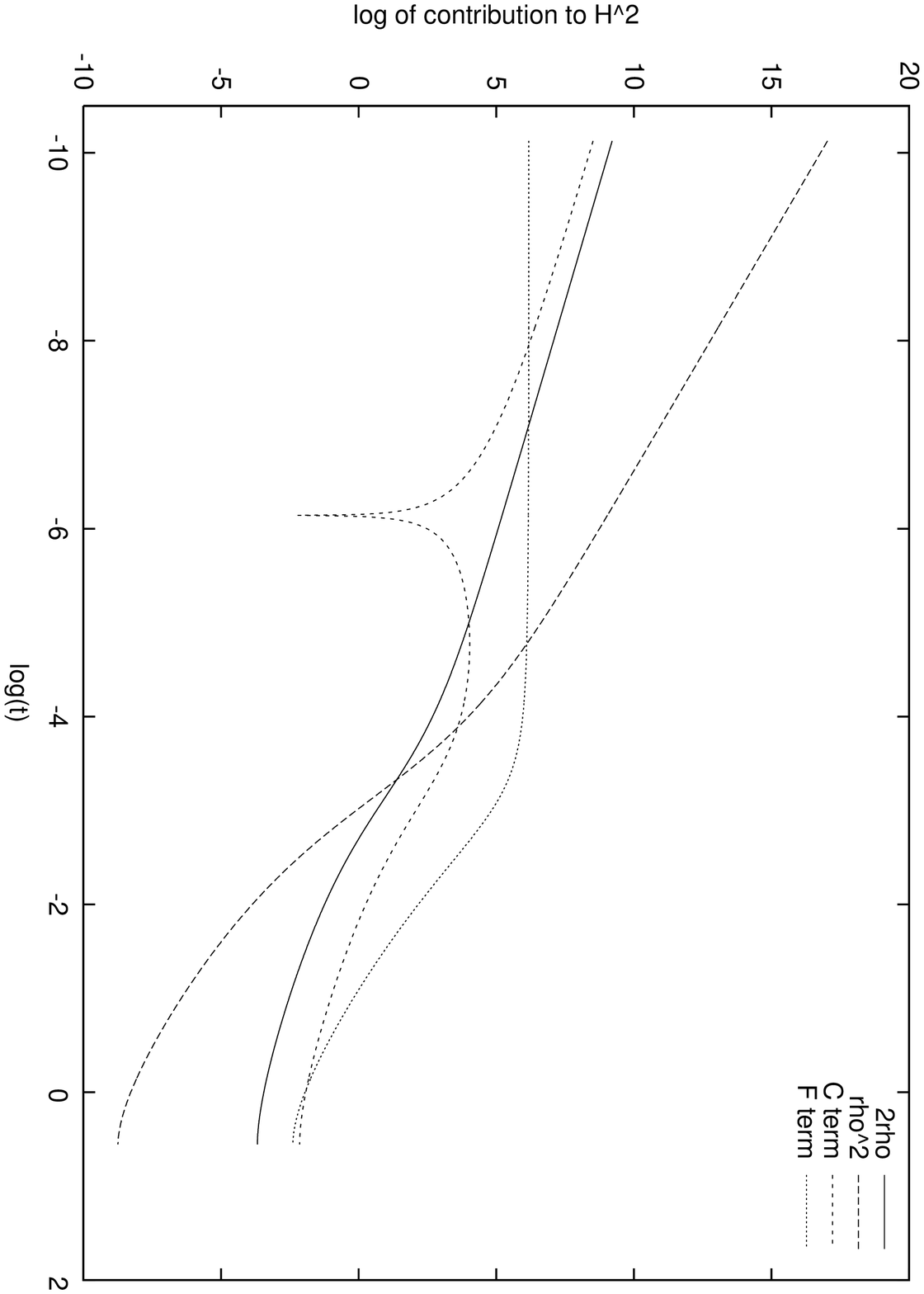,angle=180,scale=1.2}}} \quad
      \subfigure[Contributions to the Friedmann equation where the $\C$ term turns negative 
then positive.]
    {\scalebox{0.3}
      {\epsfig{file=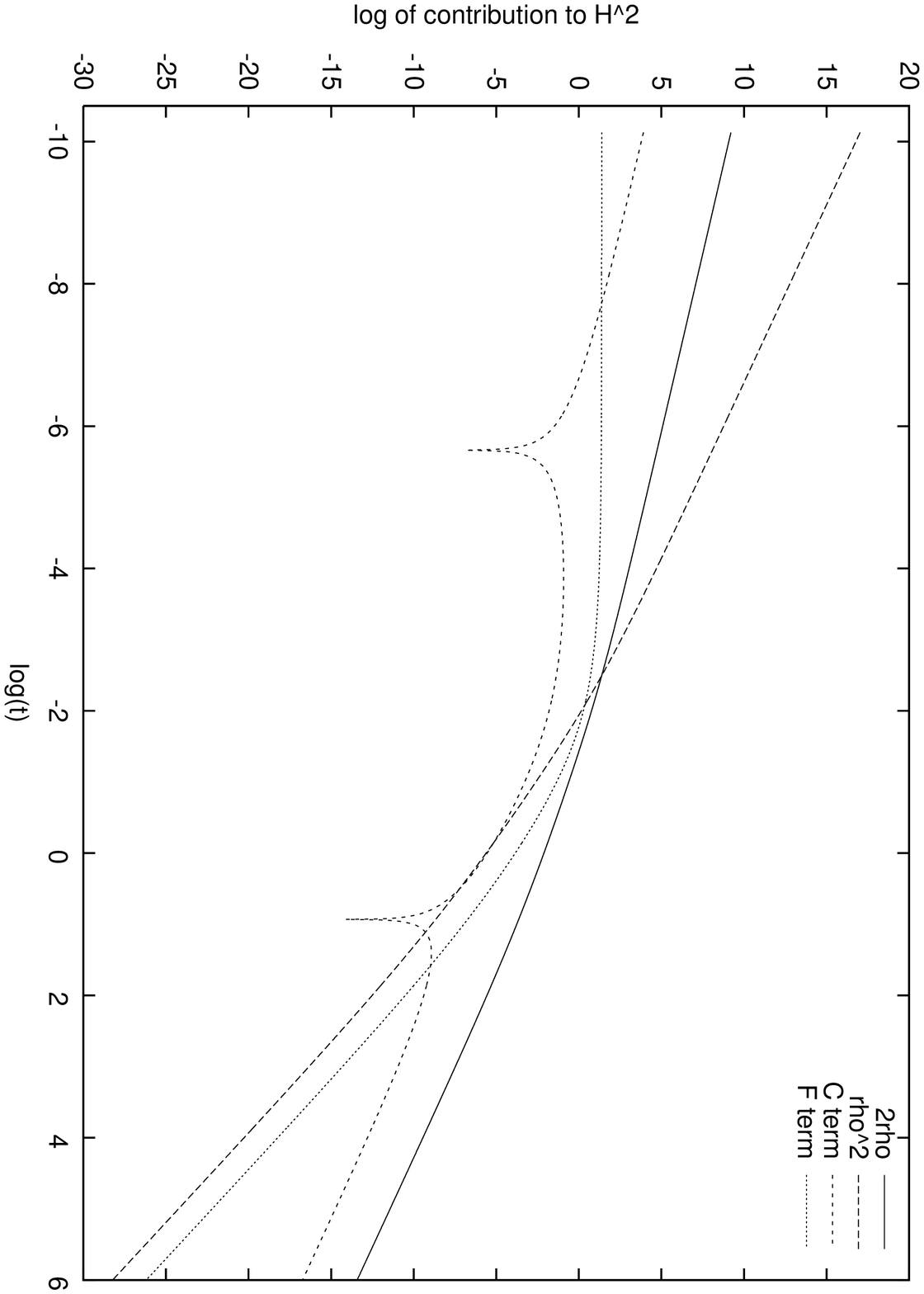,angle=180,scale=1.2}}}
      }
    \mbox{
      \subfigure[Graph of $(\C/r^4)/(2\rho)(t \rightarrow \infty)$ for 
$\alpha= 1.0, 0.4, 0.2, 0.02$ against different initial values for the $F$ term.]
    {\scalebox{0.3}
      {\epsfig{file=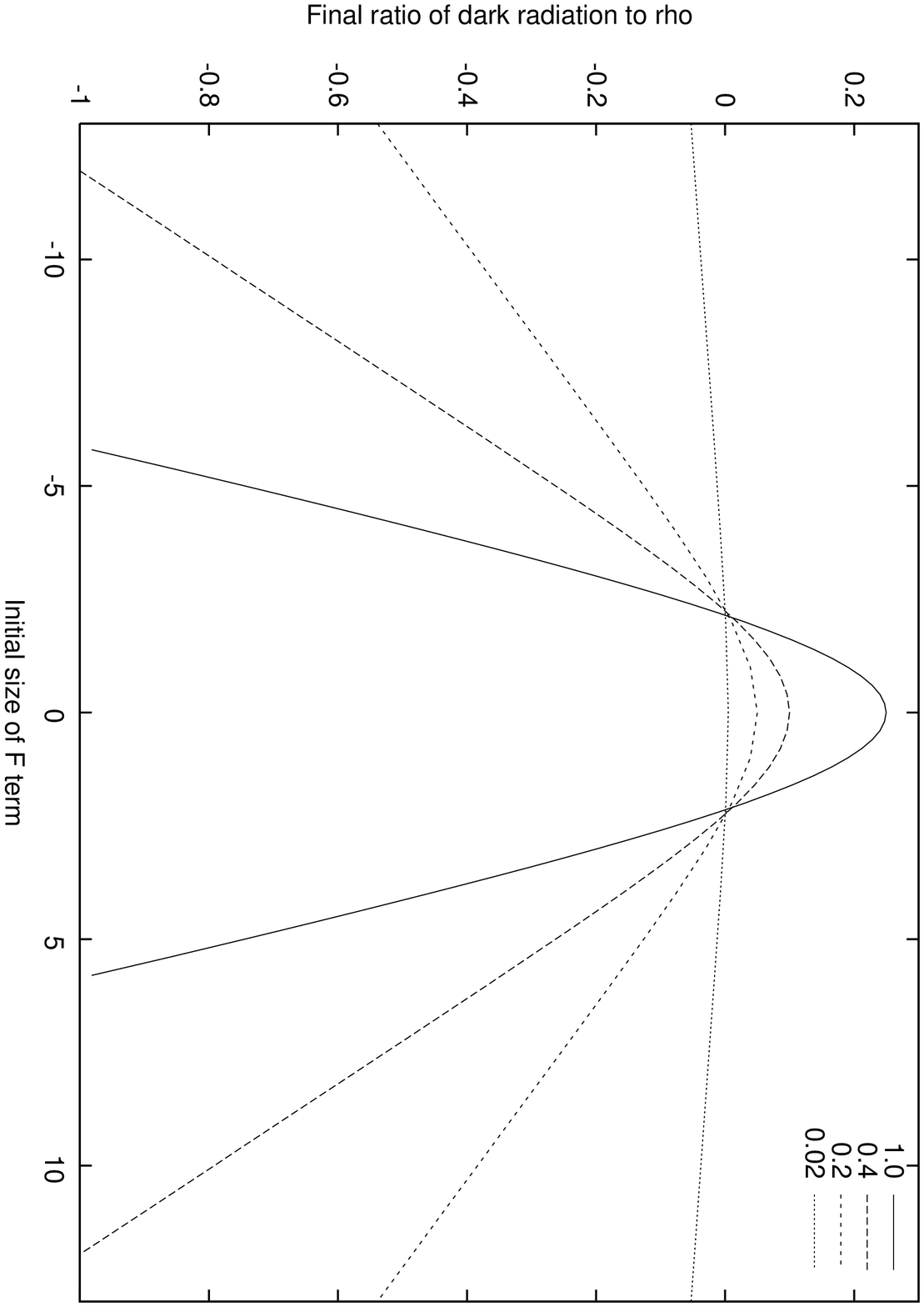,angle=180,scale=1.2}}} \quad
      \subfigure[Graph of $(\C/r^4)/(2\rho)(t \rightarrow \infty)$ against 
initial $F$ term for $\alpha=0.02$ and $\C_i= 0.5, 0.25, 0.0, -0.25$. 
Note the nucleosynthesis bounds.]
    {\scalebox{0.3}
      {\epsfig{file=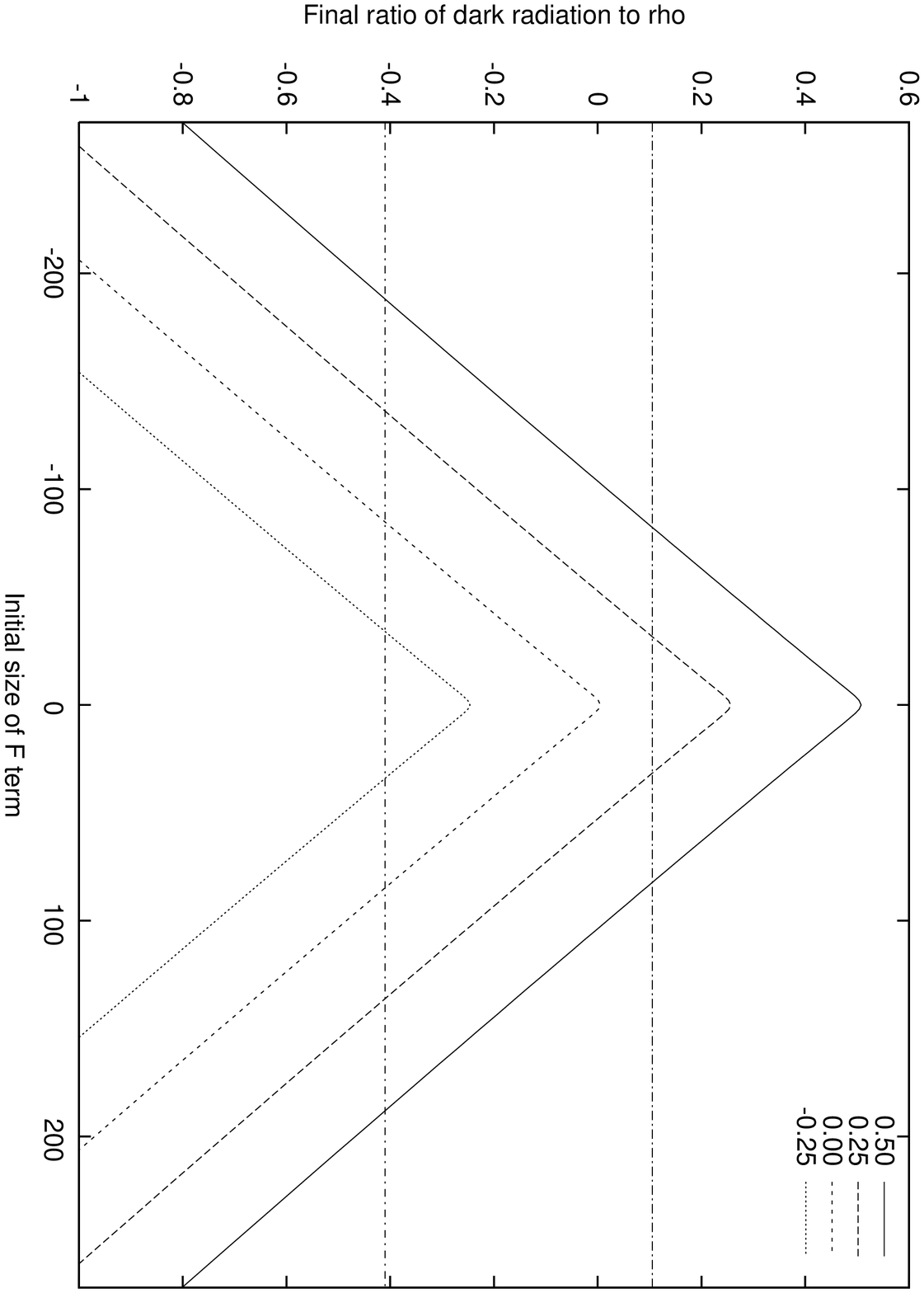,angle=180,scale=1.2}}} 
      }
    \caption{Numerical Results}\label{gra2}
  \end{center}
\end{figure}

Previously we have studied some aspects of non-$Z_2$ symmetric graviton-emitting branes in certain 
limits, but in order to gain a full understanding of the effects of non-$Z_2$ symmetry in this scenario 
we solved the equations~(\ref{dimeq}), (\ref{r}), (\ref{C}) and (\ref{F}) 
numerically for a range of values of both $\C$ and $F$ and the 
results are shown in figures \ref{gra1} and \ref{gra2}.

Figure~\ref{gra1}(a) shows the time dependence of the four contributions to the Friedmann equation: 
$\rho^2$, $2\rho$, $\C/r^4$ and $F^2/((\rho+1)^2 r^8)$, on a $\log$ scale, with the initial values 
(labelled by a subscript $i$) 
taken to be $2\rho_i = 10000$, $\C_i/r_i^4=500$ and $F_i/((\rho_i+1) r_i^4)=0.05$. The initial time 
$t_i$ was chosen to be in agreement with equation~(\ref{it1}) such that $t_i = ((4+\alpha)\rho_i)^{-1}$ 
in order for $t=0$ to be the time of the initial singularity.
The first thing to notice 
is the behaviour of each of the terms as $t\rightarrow 0$. One can see that the $\rho^2$ term dominates
as expected, the $F$ term tends to a constant in agreement with equation~(\ref{ft}), and the 
$\C$ term is proportional to $\rho$ (as predicted by equation~(\ref{ct})) and does not necessarily 
tend to a constant as assumed by~\cite{lang}. As $t \rightarrow \infty$, the brane cools and graviton 
emission is hence reduced. Therefore, as in the non-emitting case, the $F$ term is proportional 
to $\rho^2$, and $\C / r^4$ is proportional to the dominant $2\rho$ term. Note that the ratio 
of dark radiation to normal radiation defined by,
\be
\epsilon(t) \;\; \equiv \;\; \frac{(\C/r^4)}{(2\rho)}, 
\ee
increases only during the transition between these two limits, hence 
numerical analysis is definitely required to determine the final dark radiation content of the brane
world model.

The evolution of $\C$, $2F$, $\C^+$ and $\C^-$ is shown in figure~\ref{gra1}(b), using the same 
initial conditions as in the previous figure. Note the behaviour of $\C^+$: initially it decreases, 
as the brane has yet to emerge from the white hole on its $+$ve side. This corresponds 
to the extended part of the Vaidya spacetime as shown in figure~\ref{Vaid1}, where any graviton 
emissions actually leave and hence reduce the size of the white hole. Once the brane passes 
the horizon on its $+$ve side, then $\C^+$ begins to increase. $\C^-$ increases from zero in a 
similar manner to the limited cases discussed by~\cite{Leeper2}. 
Figure~\ref{gra1}(c) shows a similar situation except the initial value of $\C$ 
(and hence $\C^-$) 
are both negative. The initial conditions are the same as the previous diagrams apart from 
now $\C_i/r_i^4 = -400$. One can see that $\C^-$ changes from negative to positive. This means that on the 
$-$ve side of the brane there is initially a naked singularity ($\C^-<0$) 
which later turns into a black hole ($\C^- >0$). This corresponds to the conformal diagram 
given in figure~\ref{Vaid0}, where the incoming gravitons always cause the mass parameter to increase.
In general, all possible behaviour of $F$ and $C$ can be explained conceptually by figures~\ref{Vaid1}
and \ref{Vaid0}. 

The evolution of $\C(t)$ for different initial values of the $F$ term (ranging from 0 to 1.5) 
are shown in figure~\ref{gra1}(d). It can be seen that at early times $\C(t)$ decreases with the 
$\C(t)\sim t^{-\alpha/(4+\alpha)}$ behaviour given by equation~(\ref{ctt}), while at late times 
when there is less graviton emission $\C(t)$ tends to a constant; the specific value of which 
is lower for larger initial $F$. One can see that the conclusions made by~\cite{Leeper2}, that 
a non-zero $F$ can help reduce the final value of the dark radiation term and hence satisfy 
the nucleosynthesis bounds, are correct for small 
initial values of $F$. However, what has not been realised previously is that if $F$ is too large 
then, as is shown in figure~\ref{gra1}(d), the final value of $\C(t)$ can be negative. This 
raises the possibility of not only violating the lower nucleosynthesis bound on $\C$, but 
also of leading to $H=0$, at which time the universe begins to collapse. An example of this 
is given in figure~\ref{gra2}(a) which has initial values $\C_i/r_i^4 =5000$ 
and $F_i/((\rho_i+1) r_i^4)= 22$, and shows the magnitude of the contributions to the Friedmann 
equation on a log scale. One can see that at a time given by $\log(t) \approx -6.1$ the $\C$ term 
changes from positive to negative, and its magnitude subsequently becomes larger than the $2\rho$ term. Since at 
late times the $F$ term decreases much faster than the $C$ term, the latter term 
will dominate. This leads to $H=0$ 
at a time given by $\log(t) \approx 0.56$, and hence the universe collapses. Note that previous 
analysis of non-$Z_2$ symmetry in the non-emitting case found that a large $F$ term was 
phenomenologically acceptable (at least in terms of the expansion rate) as this term is rapidly 
damped at late times. However, now we can see that this is not the case if the brane emits 
gravitons, as a large $F$ term leads to the universe collapsing. Figure~\ref{gra2}(b) shows the 
case where the non-$Z_2$ symmetry actually helps satisfy the nucleosynthesis bounds. 
Initially $\C_i/r_i^4 = 50$ and $F_i/((\rho_i+1) r_i^4)= 2$ and here the dark radiation term 
changes from positive to negative and then back to positive again, and the final ratio $\epsilon_f$
is given by, 
\be
\epsilon_f = \epsilon(t\rightarrow \infty) < 0.037,
\ee 
which is well inside the current constraints.
For convenience we now define the value of the $F$ term as $t\rightarrow 0$ as,
\be
F^{(T)}_i \;\; = \;\; \left. \frac{F}{(\rho + 1)r^4} \right|_{t \rightarrow 0}.
\ee
Note that both $F^{(T)}_i$ and $\epsilon_f$ have well defined constant values.

In order to understand the full effect of non-$Z_2$ symmetry in this scenario, 
we graphed the dependence of the 
final ratio $\epsilon_f$ on the initial size of the $F$ term given by $F^{(T)}_i$ 
for values of the emission parameter $\alpha =$ 0.02, 0.2, 0.4, and 1.0 
as is shown in figure~\ref{gra2}(c). In all cases the initial 
size of the dark radiation parameter $\C$ was zero. When $F^{(T)}_i$ is small, the larger 
$\alpha$ is the larger $\epsilon_f$ will be, as expected. However, one can see that as discussed 
above the larger the 
magnitude of $F^{(T)}_i$, the smaller the final ratio $\epsilon_f$. If $\epsilon_f < -1$ this will 
inevitably lead to $H=0$ at some point in the late time regime (when only the $2\rho$ and $\C$ 
terms can dominate) and hence cause the universe to collapse. Note how a larger $\alpha$ leads to a 
much smaller range of allowed values for $F^{(T)}_i$ that ensure that ultimately $\epsilon_f > -1$.

Figure~\ref{gra2}(d) again shows a graph of the dependence of the 
final ratio $\epsilon_f$ on the initial size of the $F$ term: $F^{(T)}_i$. However, now 
the emission parameter is set
to $\alpha = 0.02$ (as calculated by~\cite{lang}) and the initial ratio 
$\epsilon_i \equiv (\C_i/r_i^4)/ (2\rho_i)$ takes 
the values -0.25, 0, 0.25 and 0.5. The two horizontal lines represent 
the upper and lower nucleosynthesis bounds on the ratio of dark to normal radiation~\cite{NC} given by,
\be\label{nc1}
-0.41 < \epsilon_f <0.105.
\ee
We can therefore obtain a combined restriction on the initial size of the $F$ and 
$C$ terms. The results in figure~\ref{gra2}(d) can be accurately approximated by the expression:
\be\label{ep}
\epsilon_f \;\; \simeq \;\; -\frac{| F_i^{(T)} |}{205} \;+\; \frac{\alpha}{4+\alpha} 
\;+\; \epsilon_i.
\ee
The first term on the right hand side represents the effect of the non-$Z_2$ symmetry, while the 
second term is the approximate increase in the ratio of dark to normal radiation for a $Z_2$ symmetric brane 
as calculated by~\cite{lang}. Combining equations (\ref{nc1}) and (\ref{ep}) gives the 
following restriction on the
magnitude of the initial $F$ term in terms of the initial ratio $\epsilon_i$ due to the 
nucleosynthesis constraints,
\be\label{fnc}
-20.5 \;+\; 205 \: \epsilon_i \;\; < \;\; |F_i^{(T)}| \;\; < \;\; 85.1 \;+\; 205 \: \epsilon_i,
\ee
where we have kept all numerical terms to three significant figures.
This shows that if the initial dark radiation term is more than $10\%$ of the normal 
radiation content of the universe, such that $\epsilon_i > 0.1$, then $0 < |F_i^{(T)}| < 106$ 
and hence the brane {\it has} to be non-$Z_2$ symmetric; otherwise the nucleosynthesis 
constraints would not be satisfied. If on the other hand $\epsilon_i < -0.415$ then the right hand side of 
equation (\ref{fnc}) is less than zero and regardless of the size of the $F$ term, the 
nucleosynthesis constraints will be violated.

\section{Conclusions}\label{conc}

We have analysed the dynamics of a non-$Z_2$ symmetric brane that emits 
gravitons at early times, using an $AdS$-Vaidya spacetime approximation. The equations governing the 
evolution of the energy density $\rho$, the scale factor $a$, the dark radiation parameter $\C$ and 
the non-$Z_2$ symmetry parameter $F$ were derived and analysed. We then discussed the structure of 
the five-dimensional $AdS$-Vaidya spacetimes that reside either side of the brane. It was shown 
how the ingoing five-dimensional $AdS$-Vaidya spacetime was incomplete, and that an extension was 
required in order to understand some of the relevant cases of interest, including when the black hole mass 
parameter on one side of the brane starts from a non-zero value. The conceptual arguments presented 
suggested that in some cases graviton emission would cause the black hole mass parameter  
to initially decrease (while 
the brane was still inside the white hole) before increasing at later times, a fact that has 
either been ignored or has caused some confusion in the literature~\cite{lang,Leeper,Leeper2}. 
The case where the black hole mass was initially less than zero was discussed, and it was 
shown how here, graviton emission would always lead to the mass parameter increasing and in
some situations becoming positive. 
These conceptual arguments were used to interpret the analytical and numerical results in the 
rest of the paper. 

In section~\ref{ana} we solved the evolution equations in the 
high energy/early time and low energy/late time limits in order to understand the 
asymptotic behaviour of the solutions. It was shown how even though the non-$Z_2$ symmetric 
parameter $F$ decreases with time, the non-$Z_2$ symmetric term in the Friedmann equation 
remains constant as $\rho \rightarrow \infty$, just like the symmetric case. Similarly, the dark radiation 
term $\C/r^4$, if initially large was found to be proportional to $\rho$. In the late time regime
it was shown how both $\C$ and $F$ rapidly tend to constants as the brane is cooler and much 
less graviton emission is occurring.

Numerical results were then presented, showing the behaviour of each of the four terms in the 
Friedmann equation for various initial values of $\C$ and $F$. Graphs of $\C$, $F$, $\C^+$ and $\C^-$ 
were also given which exhibited the behaviour described in section~\ref{ev} 
(and in figures~\ref{Vaid1} and \ref{Vaid0}), which corresponds 
to nonzero initial values of $\C^+$ and $\C^-$: cases which have previously been ignored 
in the literature\footnote{Note that in~\cite{Leeper2} they do examine one case where 
initially $\C=0$ and $F>0$, and hence $\C^+>0$ and $\C^-<0$. The brane therefore starts inside 
the horizon on its positive side, even though the authors have assumed throughout that the brane is 
always outside the horizon.}. 
We demonstrated that in general a large non-$Z_2$ symmetric term will lead to a decrease in the 
asymptotic ratio of dark radiation to normal radiation given by $\epsilon_f$. If $F$ 
is large enough then $\C$ and hence $\epsilon_f$ 
will become negative which in some cases leads to $H=0$ at which time 
the universe will collapse. We then used the nucleosynthesis constraints on $\epsilon_f$ to restrict 
the possible initial values of the dark radiation and non-$Z_2$ symmetric terms, finding among other 
things that if the initial ratio $\epsilon_i > 0.1$ then $F$ must be nonzero and the brane has to be 
non-$Z_2$ symmetric otherwise the nucleosynthesis bound would be violated. Interestingly, these constraints 
are all independent of $M_5$, the five-dimensional Plank mass.

Another interesting mechanism, besides graviton emission by the brane, that would affect the evolution of the 
non-$Z_2$ symmetric and dark radiation terms is the possibility of the bulk black holes Hawking radiating. 
This has non-trivial effects on the evolution of $\C$ and $F$, especially when combined with graviton 
emissions, and will be investigated in~\cite{ver6}.

\vspace{0.3cm}

{\bf Acknowledgements} We are grateful to A.C. Davis and C. van de Bruck for reading and commenting on 
the manuscript, and to J. Gray and L. Patino for useful discussions. This work is supported by PPARC.

\vspace{-0.4cm}


\bibliographystyle{JHEP} \bibliography{thesis1}

\end{document}